\documentclass[prl,twocolumn]{revtex4-1}
\usepackage{amsmath,amsfonts, amssymb, amsthm, dsfont}
\usepackage{yfonts}
\usepackage{bm}
\usepackage{mathrsfs}
\usepackage{graphicx}
\usepackage{verbatim}
\usepackage{hyperref}
\usepackage{tikz}
	\usetikzlibrary{calc}

\newcommand{\di}{\mathrm{d}}

\renewcommand{\vec}[1]{{\mathbf #1}}

\renewcommand{\vr}{{\vec{r}}}
\newcommand{\vk}{{\vec{k}}}

\newcommand{\comments}[1]{}
\newcommand{\mb}[1]{\mathbf{#1}}
\newcommand{\Ref}[1]{Ref. [\onlinecite{#1}]}

\newcommand{\spd}[1]{(#1+1)D}
\newcommand{\ftj}{\ensuremath{\mathsf{T}}} 
\newcommand{\zr}{\mathbb{Z}_2^{\mathsf{R}}}

\newcommand{\mysection}[1]{\emph{#1} -- }

\def\sp{\mathfrak{sp}}
\def\so{\mathfrak{so}}

\makeatletter
\def\l@subsubsection#1#2{}
\makeatother

\usetikzlibrary{decorations.pathreplacing,decorations.markings}
\tikzset{middlearrow/.style={
        decoration={markings,
            mark= at position 0.55 with {\arrow{#1}} ,
        },
        postaction={decorate}
    }
}
\begin{document}

\title{Microscopic Theory of Surface Topological Order for Topological Crystalline Superconductors}

\author{Meng Cheng}
\affiliation{Department of Physics, Yale University, New Haven, CT 06511-8499, USA}
\date{\today}
\begin{abstract}
	We construct microscopic Hamiltonians for symmetry-preserving topologically ordered states on the surface of topological crystalline superconductors, protected by a $\mathbb{Z}_2$ reflection symmetry. Starting from $\nu$ Majorana cones on the surface,  we show that the semion-fermion topological order emerges for $\nu=2$, and more generally, $\mathrm{SO}(\nu)_\nu$ topological order for all $\nu\geq 2$ and $\mathrm{Sp}(n)_n$ for $\nu=2n$ when $n\geq 2$.
\end{abstract}

\maketitle

\mysection{Introduction}
Three-dimensional topological insulators and superconductors~\cite{Qi_RMP2011, Hasan_RMP2010}, prominent examples of symmetry-protected topological (SPT) phases, feature topologically protected gapless surface states. It has been established that the nontrivial nature of the bulk phase manifests in the quantum anomalies of the boundary states: it is impossible to realize the same kind of physics in truly two-dimensional systems with the given symmetries. Recently, investigations of interaction effects reveals a new type of surface termination~\cite{vishwanath2013}, namely the surface may enter a fully gapped topologically-ordered state, known as surface topological order (STO), while preserving all symmetries (i.e. they are symmetry-enriched topological (SET) phases), and the anomaly manifests in the way that surface anyon excitations transform under the symmetry~\cite{barkeshli2014SDG, Chen2014}.  

Despite the rapid progress in the theoretical understanding of STOs and anomalous SETs in general, the question of microscopic realizations of STOs, in particular starting from non-interacting surface states, remains largely open.  For time-reversal-invariant class AII topological insulators (TI) and class DIII superconductors (TSC),  several different approaches, including vortex condensation~\cite{wang2013b, ChongWangPRB2014, MetlitskiPRB2015, metlitski2014, Seiberg2016, witten2016, NLSM, QiPRL2015}, coupled layer constructions~\cite{wang2013, chen2014b} and decorated Walker-Wang models~\cite{BurnellPRB2014, Fidkowski13, chen2014b, Chen2014} have been pursued and a number of possible STOs have been identified, leading to a rather complete understanding of the classifications of interacting TIs and TSCs in \spd{3}~\cite{ChongWangPRB2014,WangScience2014, metlitski2015}. However, these methods do not provide much clue about what kind of microscopic interactions on the Dirac/Majorana surface can stablize the STOs. Besides, the question of what type of STO can appear on the surface of TSCs with an odd number of Majorana cones has remained out of reach for the aforementioned approaches.

In this work we set out to answer these questions for topological crystalline superconductors (TCSC), protected by a $\mathbb{Z}_2$ reflection symmetry group (denoted by $\zr=\{\mathds{1}, \mathsf{R}\}$)~\cite{ChiuPRB2013, MorimotoPRB2013, ShiozakiPRB2014}. Although time-reversal and mirror reflection are different symmetries in condensed matter systems, from the point of view of low-energy theory often with emergent Lorentz symmetry they can be treated on equal footing~\cite{kapustin2014, cho2015, Kapustin2015b, ThorngrenElse}, and indeed the bulk classification and surface states of $\zr$ TCSC are very similar to those of class DIII TSC. With non-interacting surface theories as the starting point, we develop a \emph{microscopic} construction of STOs, employing the powerful technique of bosonization.  We explicitly construct microscopic Hamiltonians to realize STOs known from previous works, and also present a family of new non-Abelian STOs -- the $\mathrm{SO}(\nu)_{\nu}$ Chern-Simons theories -- for TCSCs having $\nu$ number of surface Majorana cones. Our method thus provides a constructive answer to the issue of STO connected to an odd number of Majorana cones. In addition, while we focus on STOs of TCSCs in the main text, our theory provides a general framework for microscopic constructions of \spd{2} topological phases enriched by reflection symmetry.

\mysection{Microscopic constructions}
Our construction is inspired by the ``dimensional reduction'' perspective on reflection SPT phases~\cite{IsobePRB2015, SongPRX2017}, relating $\zr$ SPTs in three dimensions to on-site $\mathbb{Z}_2$ SPTs in two dimensions. 
To make the idea concrete, consider a non-interacting TCSC protected by the $\zr$ symmetry. We will set the mirror plane to the $x=0$ plane. An integer topological invariant $\nu$ of the TCSC bulk determines the number of Majorana cones on a reflection-symmetric surface~\footnote{More precisely, $\nu$ determines the difference between Majorana cones of opposite chiralities.}, say the $xy$ plane $z=0$. We describe the Majorana cones with the following Hamiltonian:
\begin{equation}
	{H}_0=\sum_{a=1}^\nu \int\di^2\vr\,\chi_a^\ftj(-i\sigma_z\partial_x +i\sigma_x\partial_y )\chi_a.
	\label{}
\end{equation}
Here $\chi$ is a two-component Majorana field and transforms under the reflection symmetry as $\mathsf{R}\chi(x,y)\mathsf{R}^{-1}=\sigma_x\chi(-x,y)$.
A uniform mass term $\int\di^2\vr\, M\chi^\mathsf{T}_a \sigma_y \chi_b$ is forbidden by the reflection symmetry. However, we can allow a spatially-varying mass $M$ and if $M$ is an odd function of $x$, i.e. $M(-x)=-M(x)$, the mass term preserves $\mathsf{R}$.
For instance, we can set $M(x)=M\,\text{sgn}(x)$ where there is a mass domain wall at $x=0$. In this case, there are $\nu$ chiral Majorana fermions localized on the domain wall~\cite{HsiehPRB2016}, which explains the $\mathbb{Z}$ classification of non-interacting TCSCs.

\begin{figure}[htpb]
	\centering
	\includegraphics[width=\columnwidth]{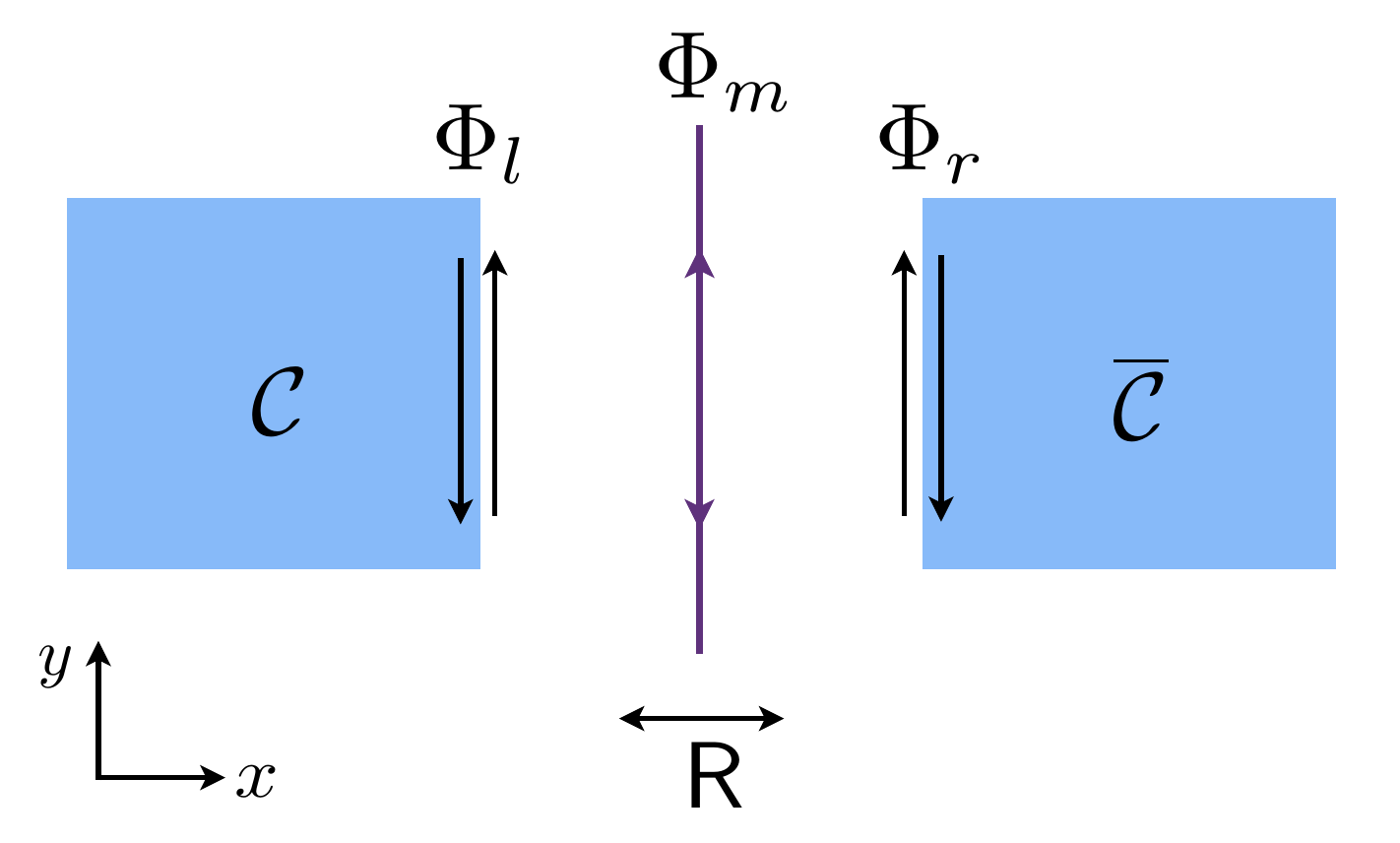}
	\caption{Illustration of the setup for the microscopic construction, from a top view of the TCSC surface. $\Phi_m$ denotes the fields on the mirror plane. $\Phi_l$ and $\Phi_r$ are edge modes of the topological phases $\mathcal{C}$ and $\bar{\mathcal{C}}$, related to each other by $\mathsf{R}$.}
	\label{fig:setup}
\end{figure}

In this setup, to obtain a fully gapped surface we need to introduce interacting topological order to gap out the remaining chiral Majorana fermions, while preserving the $\mathsf{R}$ symmetry. We attach a suitable \spd{2} topological phase $\mathcal{C}$ to the $x<0$ half plane, and its mirror image $\overline{\mathcal{C}}$ to the other half $x>0$, so the $\mathsf{R}$ symmetry is respected microscopically. The low-energy theory consists of the edge theories of $\mathcal{C}$ and $\overline{\mathcal{C}}$, as well as the gapless channel on the mirror axis:
\begin{equation}
	\mathcal{L}=\mathcal{L}_\text{edge}[\Phi_l] + \mathcal{L}_\text{edge}[\Phi_r] + \mathcal{L}_m[\Phi_m].
	\label{}
\end{equation}
Here we use $\Phi_{l/r}, \Phi_m$ to denote the corresponding field variables, as illustrated in Fig. \ref{fig:setup}. The reflection symmetry swaps $\mathcal{C}$ with $\overline{\mathcal{C}}$, therefore
\begin{equation}
	\mathsf{R}: \Phi_l \leftrightarrow \Phi_r.
	\label{}
\end{equation}
To construct a STO, we should be able to find $\mathsf{R}$-symmetry preserving interactions to gap out all the low-energy modes, which is the subject of the rest of the paper. We will also analyze how the $\mathsf{R}$ symmetry acts in the resulting STOs.

\begin{figure}[t!]
	\centering
	\includegraphics[width=0.8\columnwidth]{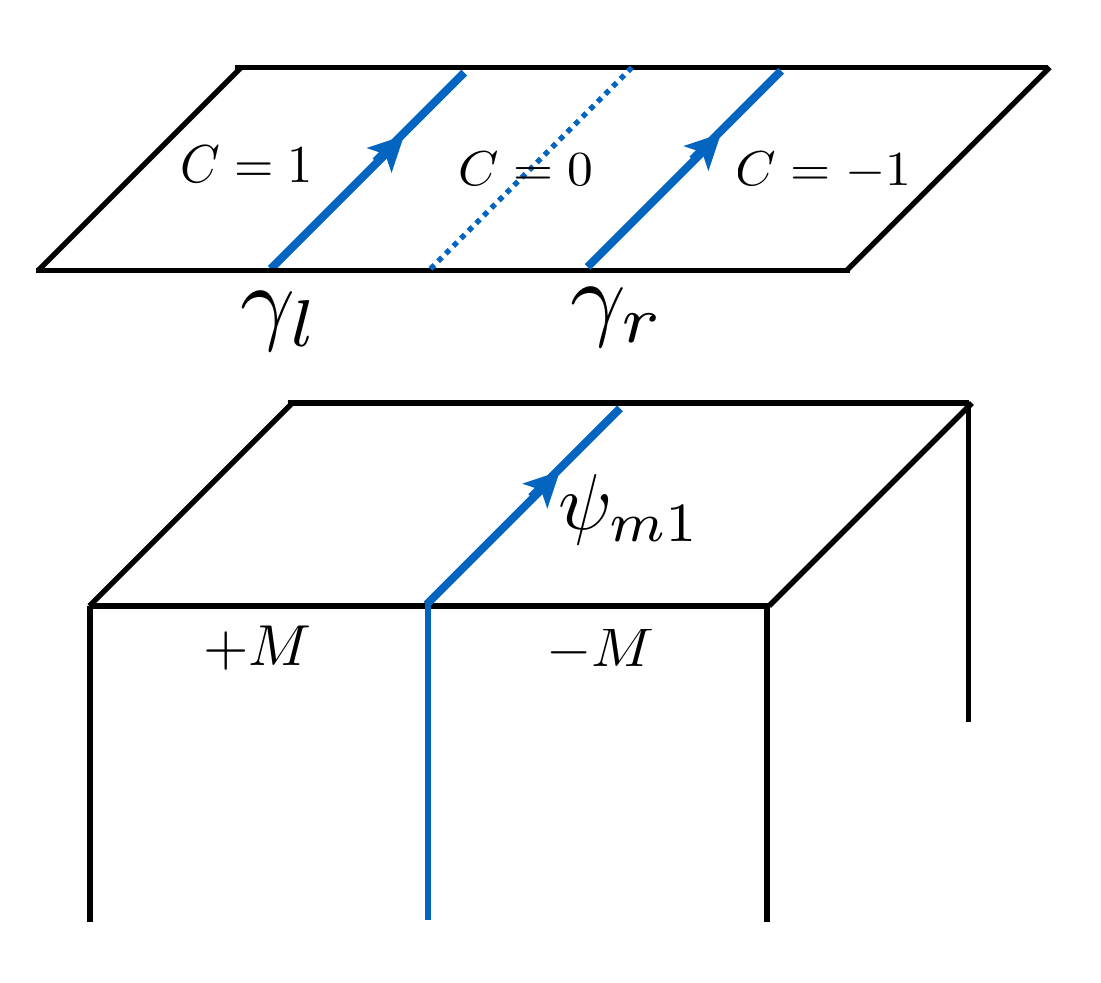}
	\caption{Illustration of the $\nu=2$ surface, with an additional layer of 2D superconductor. Here $C$ is the Chern number of the superconducting state.}
	\label{fig:recon}
\end{figure}

\mysection{$\nu=2$: Semion-fermion theory}
We first illustrate our method for $\nu=2$ TCSCs. In the domain wall setup, there are two chiral Majorana fermions, or equivalently, a single chiral Dirac fermion $\psi_{m1}$. To faciliate the construction, we introduce an additional pair of chiral Majorana fermions $\gamma_l$ and $\gamma_r$, propagating in the \emph{same} direction as the ones on the domain wall. This pair of Majorana fermions comes from surface reconstruction, i.e. by attaching a layer of 2D topological superconductor, as illustrated in Fig. \ref{fig:recon}. We then form their linear combinations 
\begin{equation}
	\psi_{m2}=\frac{e^{\frac{i\pi}{4}}\gamma_l + e^{-\frac{i\pi}{4}}\gamma_r}{2}.
	\label{}
\end{equation}
Under the reflection the new fields transform as
\begin{equation}
	\mathsf{R}\psi_{m1}\mathsf{R}^{-1}=\psi_{m1}, \mathsf{R}\psi_{m2}\mathsf{R}^{-1}=\psi_{m2}^\dag.
	\label{}
\end{equation}
We then bosonize $\psi_{m1}\sim e^{i\phi_{m1}}, \psi_{m2}\sim e^{i\phi_{m2}}$~\cite{giamarchi2003}, and
\begin{equation}
\mathsf{R}\phi_{m1}\mathsf{R}^{-1}=\phi_{m1}, \mathsf{R}\phi_{m2}\mathsf{R}^{-1}=-\phi_{m2}.
	\label{}
\end{equation}

Now we show how to gap out the surface by introducing $\mathcal{C}=\overline{\mathrm{U}(1)_2}$ topological order (i.e. the image of a $\mathrm{U}(1)$ Chern-Simons theory at level $2$). Edge states of a $\mathrm{U}(1)_{\pm 2}$ state can be described by a chiral Luttinger liquid~\cite{wen04}. Together with the bosonized chiral fermions on the domain wall, we denote the fields collectively as $\Phi=(\phi_l, \phi_r, \phi_{m1}, \phi_{m2})^\mathsf{T}$, where $\phi_l$ and $\phi_r$ are the chiral bosons for the $\mathrm{U}(1)_{\mp 2}$ states on the two sides, propagating in the opposite direction to the domain wall fermions.
The Luttinger liquid Lagrangian density reads
\begin{equation}
	\mathcal{L}_\text{LL}=\frac{1}{4\pi}\partial_t \Phi^\ftj \mathcal{K}\partial_y \Phi - \partial_y\Phi^\ftj \mathcal{V} \partial_y\Phi.
	\label{}
\end{equation}
Here $\mathcal{K}$ is the following symmetric integer matrix: 
\begin{equation}
	\mathcal{K}=
	\begin{pmatrix}
		-2 & 0 & 0 & 0\\
		0 & -2 & 0 & 0\\
		0 & 0 & 1 & 0\\
		0 & 0 & 0 & 1\\
	\end{pmatrix},
	\label{}
\end{equation}
and $\mathcal{V}$ is the velocity matrix.

To gap out the edge and domain wall modes, we add the following backscattering potentials:
\begin{equation}
	\mathcal{L}'=u\cos (2\phi_l + \phi_{m1}-\phi_{m2}) +u \cos (2\phi_r + \phi_{m1}+\phi_{m2}).
	\label{}
\end{equation}
We assume $u<0$ for simplicity.  Notice that both terms are legitimate local interactions built out of local operators (i.e. $e^{\pm 2i\phi_{l/r}}$), and manifestly symmetric under $\mathsf{R}$. We will assume that the density-density interactions are tuned to make both consine terms relevant.  It is straightforward to check that the arguments of the cosine terms commute with each other, therefore can be pinned to classical minima of the cosine potentials simutaneously. Using the results developed in \Ref{LevinPRB2012} and \Ref{WangPRB2013}, one can show that there is a unique ground state of the cosine potentials, ruling out the possibility of spontaneous symmetry breaking.

The STO we constructed is known as the ``semion-fermion'' theory~\cite{Fidkowski13}, and was derived on the surface of the $\nu=2$ class DIII TSC by vortex condensation~\cite{Chen2014, ChongWangPRB2014}. Anyons in the theory are denoted by $\{1,s\}\times\{1,f\}$, where $s$ represents a semion and $f$ is the physical electron. The time-reversal or reflection symmetry has a unique action on the anyons: under $\mathsf{R}$, a semion $s$ has to become $sf$ since time-reversal/reflection symmetries reverse orientations.  

It is illustrative to explicitly derive the $\mathsf{R}$ action from the microscopic construction. To this end we first determine the form of anyon string operators. For concretness, imagine making two relatively small holes at $(-x_0,0)$ and $(x_0,0)$. Excitations on these two holes can also be described by $\mathrm{U}(1)_{\mp 2}$ chiral Luttinger liquid theories, whose fields are denoted by $\phi_l'$ and $\phi_r'$ respectively. Define the following string operator
\begin{equation}
	W_s= e^{i(\phi_l'-\phi_l-\phi_r-\phi_{m1}+\phi_r')},
	\label{}
\end{equation}
which creates $s$ and $sf$ at the two holes respectively, with an additional fermion $f$ in the middle. Because $\phi_l+\phi_r+\phi_{m1}$ is pinned by the gapping potentials, no other excitations are created except at the two holes. Apparently the string operator is invariant under $\mathsf{R}$, which is only possible when $\mathsf{R}(s)=sf$. For such anyons, we can define the reflection quantum number $\eta_s$ of the state created by applying $W_s$ to the ground state~\cite{zaletel2015, QiPRB2015, Barkeshli2016}. In the present case it is obvious that $\eta_s=1$. We discuss a modified construction with $\eta_s=-1$ in the Supplementary Material.

\mysection{$\nu\geq 3$: $\mathrm{SO}(\nu)_\nu$ Theory}
We now describe a general construction of STOs for $\nu\geq 2$ TCSCs. In order to handle odd $\nu$, we will use non-Abelian bosonization~\cite{Witten84, difrancesco}, which we now briefly review. The field theory of $n$ chiral Majorana fermions enjoys an emergent $\mathrm{SO}(n)$ symmetry, associated with the following conserved currents:
\begin{equation}
	J^\beta(z)=\frac{i}{2}\sum_{ab}\chi_{a}(z)\Sigma_{{ab}}^\beta \chi_{b}(z).
	\label{}
\end{equation}
Here $z=\tau+iy$ is the complex space-time parameter. The indices ${a,b}$ run from $1$ to $n$. $\Sigma^\beta$ are antisymmetric $n\times n$ matrices which generate the $\mathfrak{so}(n)$ Lie algebra. A common choice of the basis is labeled by $(r,s)$ where $1\leq r<s\leq n$, and the corresponding matrix reads:
	$\Sigma^{(r,s)}_{ab}=\delta^{r}_a\delta^s_b-\delta^r_b\delta^s_a$.
The currents satisfy the following operator product expansion:
\begin{equation}
	J^\alpha(z)J^\beta(w)=\frac{\delta_{\alpha\beta}}{(z-w)^2}+\frac{if^{\alpha\beta\gamma}J^\gamma(w)}{z-w}+\dots,
	\label{}
\end{equation}
where $f_{\alpha\beta\gamma}$ are the structure constants of $\mathfrak{so}(n)$ Lie algebra.
With the definition of currents, the energy-momentum tensor of the free theory can be built using the Sugawara construction:
\begin{equation}
	T(z)=\frac{1}{2(n-1)}\mb{J}(z)\cdot\mb{J}(z).
	\label{}
\end{equation}

The central idea of our construction is to exploit the following conformal embedding of Wess-Zumino-Witten models~\cite{difrancesco, SahooPRB2016}:
\begin{equation}
	\mathfrak{so}(\nu)_\nu\times\mathfrak{so}(\nu)_\nu \subseteq \mathfrak{so}(\nu^2)_1.
	\label{}
\end{equation}
This can also be viewed as a level-rank duality of the corresponding Chern-Simons gauge theories, which implies that the $\mathrm{SO}(\nu)_\nu$ theories are time-reversal/mirror symmetric, first observed in \Ref{Aharony2016}. 
The conformal embedding stems from the natural embedding of the $\mathfrak{so}(\nu)$ matrix Lie algebra to $\mathfrak{so}(\nu^2)$. To see the embedding explicitly, we represent the indices for $\mathfrak{so}(\nu^2)$ matrices as a pair $\mb{a}=(a_1,a_2)$ where $a_1, a_2=1,2,\dots,\nu$. Then any $\mathfrak{so}(\nu)$ matrix can be embedded into $\mathfrak{so}(\nu^2)$ by taking tensor product with the $\nu\times\nu$ identity matrix from the left and the right. More explicitly, we define
\begin{equation}
	\bm{\Sigma}^l_{\mb{ab}}=\bm{\Sigma}_{a_1b_1}\delta_{a_2b_2}, \bm{\Sigma}^r_{\mb{ab}}=\delta_{a_1b_1}\bm{\Sigma}_{a_2b_2}. 
	\label{}
\end{equation}

We define the two $\mathfrak{so}(\nu)_\nu^{l/r}$ currents:
\begin{equation}
	\mb{J}_{l/r}(z)=\frac{i}{2}\chi(z)^\ftj \bm{\Sigma}^{l/r} \chi(z).
	\label{}
\end{equation}
It is straightforward to check that
\begin{equation}
	J_{\lambda}^\alpha(z)J_{\lambda}^\beta(w)=\frac{\nu\delta_{\alpha\beta}}{(z-w)^2}+\frac{if_{\alpha\beta\gamma}J_{\lambda}^\gamma(w)}{z-w}+\dots, \lambda=l,r
	\label{}
\end{equation}
and $J_l^\alpha(z)J_r^\beta(w)$ contain no singular terms, for $\alpha,\beta$ valued in the indices $\{(r,s), 1\leq r<s\leq \nu\}$. This is the expected current algebra for two commuting copies of $\mathfrak{so}(\nu)_\nu$. Furthermore, we can also check that the Sugawara energy-momentum tensors decompose:
\begin{equation}
	T_{\mathfrak{so}(\nu^2)_1}=T_{\mathfrak{so}(\nu)_\nu^l} + T_{\mathfrak{so}(\nu)_\nu^r}.
	\label{}
\end{equation}
We provide details of these calculation in the Supplementary Material.

We now apply the decomposition of the current algebra to the present problem. We introduce additional $\nu(\nu-1)$ chiral Majorana fermions (with the same chirality) by attaching $p_x\pm ip_y$ superconductors to the surface in a reflection-symmetric manner. We index the $\nu^2$ modes by $(a_1, a_2)$, such that the reflection action is given by
\begin{equation}
	\mathsf{R}\chi_{(a_1,a_2)}\mathsf{R}^{-1}=\chi_{(a_2,a_1)}.
	\label{}
\end{equation}
All $\chi_{(a_1,a_2)}$ with $a_1\neq a_2$ are thus ``trivial'', coming from surface reconstruction (i.e. attaching $p_x+ip_y$ layers). Only the $\nu$ ``diagonal'' ones $\chi_{(a,a)}$ are invariant under reflection. In other words, the bulk topological invariant is still $\nu$. Importantly, we find that under reflection:
\begin{equation}
	\mathsf{R}\mathbf{J}_l\mathsf{R}^{-1}=\mathbf{J}_r.
	\label{eqn:R_currents}
\end{equation}
This fact is the key for the construction to work.

The conformal embedding naturally motivates $\mathrm{SO}(\nu)_\nu$ as the STO. Since the $\mathrm{SO}(\nu)_\nu$ topological order is relatively unfamiliar, we here give a coupled wire construction the STO~\cite{teo2011}, generalizing that of $\mathrm{SO}(3)_3$ given in Ref. [\onlinecite{SahooPRB2016}]. Let us make an array of domain walls:
\begin{equation}
	M(x)=
	\begin{cases}
		-M_0 & (2k-1)L < x < 2kL\\
		M_0 & 2kL < x < (2k+1)L
	\end{cases}, k\in\mathbb{Z}.
	\label{}
\end{equation}
At each domain wall $x=jL$, there is a bundle of $\nu$ chiral Majorana fermions, labeled as $\chi_{j,a}$ with $a=1,2,\dots,\nu$. For even/odd $j$, the fermions move upwards/downwards. We then attach to each domain a 2D topological superconductor with Chern number equal to $\pm \frac{\nu(\nu-1)}{2}$, such that there are exactly $\nu^2$ chiral Majorana fields at each domain wall. For each bundle we can use the splitting to define $\mathfrak{so}(n)_n$ currents $\mathbf{J}_{l/r, j}$. Clearly the setup can be made reflection-symmetric, with the following $\mathsf{R}$ transformations:
\begin{equation}
	\begin{gathered}
		\mathsf{R}\chi_{j,(a_1,a_2)}\mathsf{R}^{-1}= \chi_{-j, (a_2,a_1)}\\
		\mathsf{R}\mb{J}_{l, j}\mathsf{R}^{-1}=\mb{J}_{r,-j}.
	\end{gathered}
	\label{eqn:Rtrans}
\end{equation}

\begin{figure}[t!]
	\centering
	\includegraphics[width=0.8\columnwidth]{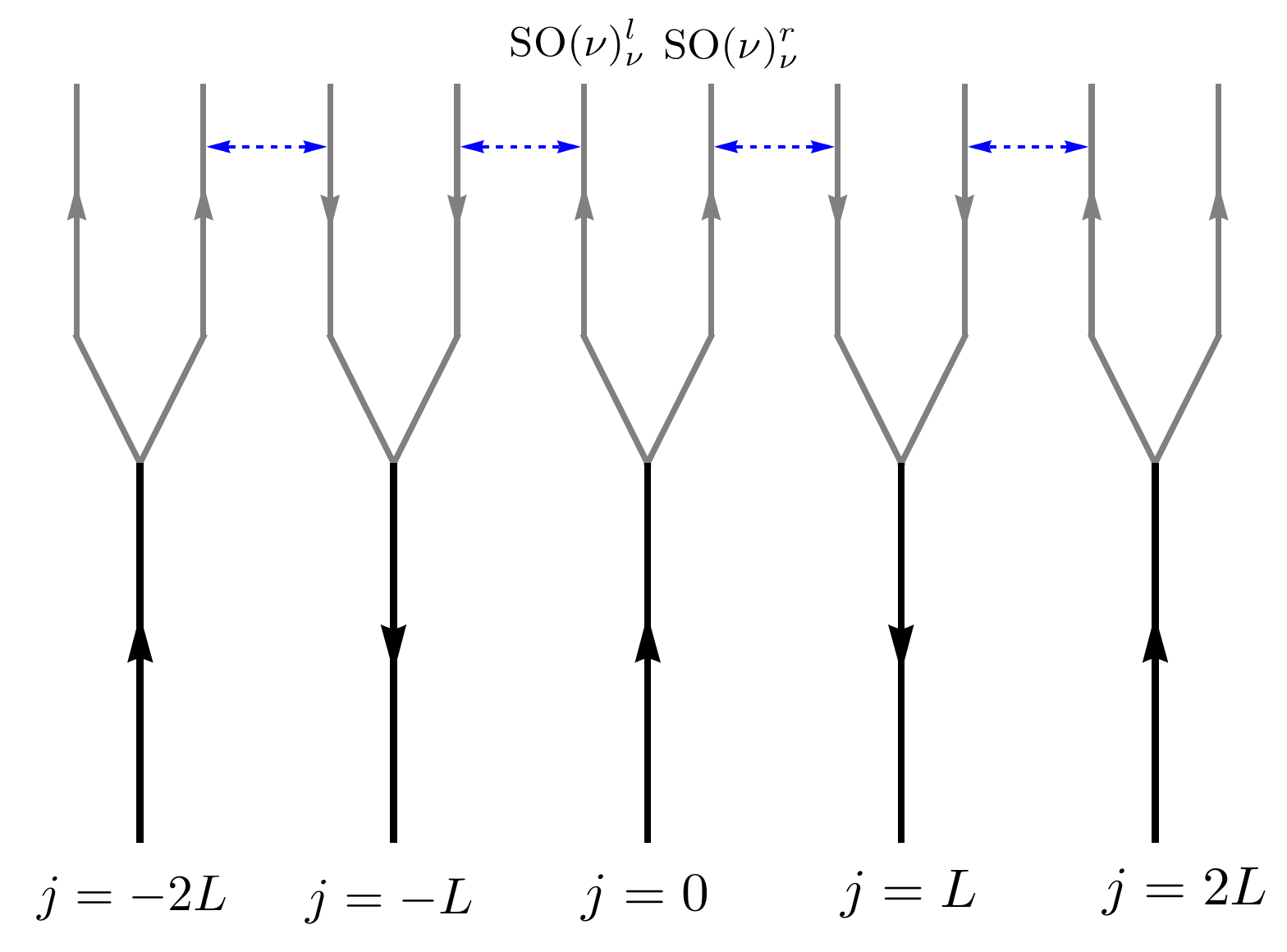}
	\caption{Wire construction of the $\mathrm{SO}(\nu)_\nu$ STO.}
	\label{fig:wire_so}
\end{figure}

We are finally prepared to write down the gapping interactions:
\begin{equation}
	\mathcal{H}'=u\sum_j \mathbf{J}_{r,j}\cdot \mathbf{J}_{l,j+1}.
	\label{eqn:H}
\end{equation}
Notice that $\mb{J}_{r,j}$ and $\mb{J}_{l,j+1}$ have opposite chiralities. See Fig. \ref{fig:wire_so} for an illustration of the construction.
It is clear that with the transformation rules given in Eq. \eqref{eqn:Rtrans} this Hamitonian preserves the reflection symmetry. $\mathcal{H}'$ consists of four-fermion interactions, and when $u>0$ the term is marginally relevant, flowing to strong coupling and gapping out the left- and right-moving degrees of freedom from neighboring domain walls~\cite{SahooPRB2016}. 

We now analyze the resulting STOs for $\nu=3,4$ more carefully:
\begin{enumerate}
	\item {$\mathrm{SO}(3)_3$}.  $\mathrm{SO}(3)_3$ as an anyon model can be described as the subcategory of integer spins in $\mathrm{SU}(2)_6$ theory~\cite{Fidkowski13}. There is only one nontrivial anyon type, denoted by $s$, with the fusion rule and topological twist given by
\begin{equation}
	\begin{gathered}
	s\times s= 1 + s + sf,\: \theta_s=i.
	\end{gathered}
	\label{}
\end{equation}
The action of the reflection symmetry is again uniquely fixed:
	$\mathsf{R}(s)= sf$.
The possibility of the $\mathrm{SO}(3)_3$ theory as the STO of a root class DIII TSC was first proposed in Ref. [\onlinecite{Fidkowski13}] ~\footnote{Notice that the theory was called $\mathrm{SO}(3)_6$ there.} Recently, based on a conjectured formula relating the data of STO to the bulk topological invariant~\cite{wang2016, Tachikawa2, TachikawaPRL2017}, it was suggested that $\mathrm{SO}(3)_3$ should be realized on the surface of $\nu=3$ TSC. Here we give a constructive proof of this result for TCSC.
\item {$\mathrm{SO}(4)_4$}. Since $\mathfrak{so}(4)\simeq \mathfrak{su}(2)\oplus \mathfrak{su}(2)$, $\mathrm{SO}(4)_4$ can be understood as
	$\mathrm{SO}(4)_4\simeq \frac{\mathrm{SU}(2)_4\times\mathrm{SU}(2)_4}{(2,2)}$.
This expression means that we condense the $\mathbb{Z}_2$ boson $(2,2)$ in $\mathrm{SU}(2)_4\times\mathrm{SU}(2)_4$. The spectrum of anyon types can be worked out using the theory of anyon condensation~\cite{moore1989, bais2009, eliens2013, NeupertPRB2016}, and we list them  in Table \ref{tab:so44}. We should notice that, unlike $\mathrm{SO}(3)_3$, $\mathrm{SO}(4)_4$ by itself is a purely bosonic theory, i.e. the physical fermions are decoupled in the topological order.

At the moment we do not know how to derive the symmetry action directly from the Hamiltonian Eq. \eqref{eqn:H} for general $\nu$. A detailed analysis of the $\mathsf{R}$ action in this SET will be presented elsewhere~\cite{elsewhere}, and we briefly summarize the results in Table \ref{tab:so44}~\footnote{Interestingly, there is in fact an ambiguity in the choice of the reflection action on anyon types. We could have chosen $\mathsf{R}: (\tfrac{1}{2},\tfrac{1}{2})\leftrightarrow (\tfrac{3}{2},\tfrac{1}{2})$, under which the theory is manifestly invariant, without involving the physical fermions at all. In other words, it is possible to define a reflection action in the bosonic topological phase $\mathrm{SO}(4)_4$ alone. However, if this is the case it means that the bulk state is a \emph{bosonic} SPT, which can not be true for $\nu=4$. Besides, such an assignment of $\mathsf{R}$ suffers from a $\mathcal{H}^3$ obstruction~\cite{elsewhere}.}.

\end{enumerate}

\begin{table}
	\centering
	\begin{tabular}{|c|c|c|c|c|c|c|c|c|}
		\hline
		$a$ & $(0,0)$ & $(2,0)$ & $(\tfrac{1}{2},\tfrac{1}{2})$ & $(\tfrac{3}{2},\tfrac{1}{2})$ & $(1,0)$ & $(0,1)$ & $(1,1)_+$ & $(1,1)_-$\\
		\hline
		$d_a$ & 1 & 1 & 3 & 3 & 2 & 2 & 2 & 2\\
		\hline
		$\theta_a$ & 1 & 1 & $i$ & $-i$ & $e^{\frac{2\pi i}{3}}$ & $e^{\frac{2\pi i}{3}}$ & $e^{\frac{2\pi i}{3}}$ & $e^{\frac{2\pi i}{3}}$\\
		\hline
		$\mathsf{R}(a)$ & $(0,0)$ & $(2,0)$ & $(\tfrac{1}{2},\tfrac{1}{2})f$ & $(\tfrac{3}{2},\tfrac{1}{2})f$ & $(1,1)_+$ & $(1,1)_-$ & $(1,0)$ & $(0,1)$\\
		\hline
	\end{tabular}
	\caption{Anyon types in $\mathrm{SO}(4)_4$, labeled by the $\mathrm{SU}(2)$ spins of their parent anyon types in $\mathrm{SU}(2)_4\times \mathrm{SU}(2)_4$ before condensation.}
	\label{tab:so44}
\end{table}

\mysection{Conclusions and discussions} 
In this work we present a microscopic theory of STOs for TCSCs. The approach makes the connection between the STOs and the non-interacting surface states very explicit. The applicability of our construction extends well beyond STOs of TCSCs. In the Supplementary Material, we discuss (i) how to construct reflection-symmetric $\mathbb{Z}_N$ toric code phases, both anomalous and non-anomalous ones (ii) a different family of non-Abelian STOs, $\mathrm{Sp}(n)_n$ CS theories, for TCSC with $\nu=2n$, and (iii) adaption of the construction to topological crystalline insulators with $\mathrm{U}(1)$ charge conservation. We have thus provided the first computation of time-reversal/parity anomalies in $\mathrm{SO}(n)_n$ and $\mathrm{Sp}(n)_n$ Chern-Simons theories~\cite{Aharony2016}.

It is useful to compare our results with previous works applying similar methods to TIs/TSCs protected by staggered time-reversal symmetry~\cite{MrossPRX2015, MrossPRL2016, MrossPRL2016b, SahooPRB2016} or glide symmetry~\cite{Lu_unpub}. There the time-reversal or reflection transformation squares to a lattice translation, and this fundamental difference in the symmetry group structure alters the classification completely for TSCs. For example, with time-reversal or reflection symmetries TSC has a $\mathbb{Z}$ classification for free fermions and $\mathbb{Z}_{16}$ for interacting ones, but for staggered time-reversal or glide symmetries even the free fermion classification is just $\mathbb{Z}_2$~\cite{SahooPRB2016, Lu_unpub}, and thus exhibit a much weaker topological protection. 


\mysection{Acknowledgement}
I am grateful to Chao-Ming Jian and Yang Qi for many illuminating conservations, Claudio Chamon and Thomas Iadecola for interesting discussions on non-Abelian bosonization, and Maissam Barkeshli for collaborations on related projects. I would like to especially acknowledge Yi-Zhuang You for explaining the embedding of $\mathfrak{sp}(n)$ into $\mathfrak{so}(4n^2)$. While the manuscript was being prepared, I learnt of a related work by S. Hong and L. Fu~\cite{Hong_toappear}.

\clearpage

\onecolumngrid

\vspace{1cm}
\begin{center}
{\bf\large Supplementary material}
\end{center}
\vspace{0.2cm}


\renewcommand{\theequation}{S\arabic{equation}}  
\setcounter{equation}{0}

\section{Summary}

In this supplementary material we study more examples of \spd{2} reflection symmetry-enriched topological phases within our model of coupled edges: (i) we construct $\mathbb{Z}_N$ toric code models with a reflection symmetry, particularly paying attention to how reflection quantum numbers can be computed in the microscopic models. (ii) we carry out the construction for STOs of topological crystalline insulators (TCI). (iii) We show that $\mathrm{Sp}(n)_n$ can arise as the STOs of topological crystalline superconductors with bulk invariant being $\nu=2n$. (iv) we also record some calculational details for the conformal embedding used in the main text.

\section{$\mathbb{Z}_N$ Toric Code}
\label{sec:toriccode}
In this section we study the example of reflection-symmetric $\mathbb{Z}_N$ toric code to illustrate the construction, in particular how the gapping potentials encode reflection symmetry fractionalization. 

In the construction, we place a $\mathbb{Z}_N$ toric code in the left half plane $x<0$, and its mirror image in the right half plane. We will also consider the possibility of an additional nonchiral Luttinger liquid on the mirror plane. We will need to couple the two edges of the $\mathbb{Z}_N$ toric code together (possibly with the LL on the mirror plane as well) to have a fully gapped phase. Notice that we can not simply gap out the two edges separately. The edges must be glued such that quasiparticles can move freely across the mirror axis.

To describe $\mathbb{Z}_N$ toric code we use a K matrix $K=N\sigma^x$, and denote the edge fields as $(\phi_l, \theta_l)^\ftj$ and $(\phi_r, \theta_r)^\ftj$ respectively. 

One possibility is to directly gap out the two edges by the following gapping potentials:
\begin{equation}
	\mathcal{L}'=-u\cos [N(\phi_l+\phi_r) - \alpha]-v\cos[N(\theta_l - \theta_r)-\beta].
	\label{}
\end{equation}
Notice that $\alpha$ can be arbitrary, but $\beta$ is fixed to be $0$ or $\pi$ by the reflection symmetry. Without loss of generality, we set $\alpha=0$ and $u, v>0$. The $\phi_l+\phi_r$ field is pinned to $\frac{2\pi n}{N}$, and $\theta_l-\theta_r$ is pinned to $\frac{2\pi n'}{N}+\frac{\beta}{N}$, where $n,n'\in \mathbb{Z}$. 

From the Hamiltonian we can determine how anyons transform under reflection. To this end we first determine the form of anyon string operators. For concretness, imagine making two relatively small holes at $(-x,0)$ and $(x,0)$. Excitations on these two holes can also be described by Luttinger edge theories, whose fields are denoted by $\Phi_l'$ and $\Phi_r'$ respectively. Define the following string operator
\begin{equation}
	W_{eL}=e^{i(\phi_l' - \phi_l)}.
	\label{}
\end{equation}
We leave the coordinate of $\phi_l'$ unspecified in the definition. Similarly we define $W_{eR}=e^{i(\phi_r'-\phi_r)}$. Joining them together, we obtain the following string operator
\begin{equation}
	W_e=W_{eL}W_{eR}= e^{i(\phi_l'-\phi_l-\phi_r+\phi_r')},
	\label{}
\end{equation}
which creates $e$ and $\bar{e}$ at the two holes. Apparently the string operator is completely invariant under reflection, which implies
\begin{equation}
	\mathsf{R}(e)= \bar{e},
	\label{}
\end{equation}
and $\eta_e=1$.

Now let us consider $m$ instead. A string to pair create $m$ and $\bar{m}$ is given by
\begin{equation}
	W_m=e^{i(\theta_l'-\theta_l+\theta_r-\theta_r')}.
	\label{}
\end{equation}
$W_m$ is obviously not symmetric under reflection, consistent with
\begin{equation}
	\mathsf{R}(m)= m.
	\label{}
\end{equation}

When $N$ is even, we expect that the string $W_m^{N/2}$ can be made reflection symmetric. To see this explicitly, we turn on gapping terms $\mathcal{L}_\text{hole}\propto -(\cos N\theta_l'+\cos N\theta_r')$ on the two holes. Under the reflection,
\begin{equation}
	\begin{split}
		\mathsf{R}W_m^{N/2} \mathsf{R}^{-1}=e^{\frac{iN}{2}(\theta_r'-\theta_r+\theta_l-\theta_l')} =e^{iN\theta_r'}e^{-iN\theta_l'} e^{iN(\theta_l-\theta_r)}W_m^{N/2}.
	\end{split}
	\label{}
\end{equation}
Projecting to the low-energy sector, we find that the reflection eigenvalue is 
\begin{equation}
	\eta_{m^{N/2}}=\langle e^{iN\theta_r'}e^{-iN\theta_l'} e^{iN(\theta_l-\theta_r)}\rangle=e^{i\beta}.
	\label{}
\end{equation}

Now assume $N$ is even and we put on the mirror axis an additional 1D Luttinger liquid, i.e. $K_m=\sigma^x$, with fields $\phi_m$ and $\theta_m$. We also impose the following $\zr$ transformation:
\begin{equation}
	\mathsf{R}\phi_m\mathsf{R}^{-1}= \phi_m+a\pi, \mathsf{R}\theta_m\mathsf{R}^{-1}=\theta_m+b\pi.
	\label{eqn:z2spt}
\end{equation}

The following perturbations can gap out the edge modes:
\begin{equation}
	\begin{split}
	\mathcal{L}'=&-u\cos [N(\phi_l+\phi_r+\phi_m)] - v\cos (N\theta_l-\theta_m) - v(-1)^b \cos (N\theta_r-\theta_m).
	\end{split}
	\label{}
\end{equation}
It is straightforward to check that these terms commute with each other. Notice that the last two terms essentially enforce $N\phi_l-N\phi_r=b\pi$ in the ground state.

We can again construct the string operators:
\begin{equation}
	\begin{gathered}
	W_e= e^{i(\phi_l'-\phi_l-\phi_r-\phi_m+\phi_r')},\\
	W_m=e^{i(\theta_l'-\theta_l+\theta_r-\theta_r')}.
	\end{gathered}
	\label{}
\end{equation}
Due to the factor $e^{-i\phi_m}$ in $W_e$, we now have $\eta_e=(-1)^a$. The computation of $\eta_{m^{N/2}}$ is essentially identical to the previous case, so $\eta_{m^{N/2}}=(-1)^b$.

When $a=b=1$, the resulting SET is anomalous. This is consistent with Eq. \eqref{eqn:z2spt} describing the edge of a $\mathbb{Z}_2$ SPT phase in two dimensions, so that the bulk is a nontrivial reflection SPT. For $N=2$, an exactly-solvable lattice model for this so-called $ePmP$ state was studied in Ref. [\onlinecite{SongPRX2017}].

\section{Semion-Fermion Theory with $\eta_s=-1$}
To get a semion-fermion STO with $\eta_s=-1$, we modify the $\mathsf{R}$ symmetry transformation on the domain wall fields:
\begin{equation}
\mathsf{R}\phi_{m1}\mathsf{R}^{-1}=\phi_{m1}+\pi, \mathsf{R}\phi_{m2}\mathsf{R}^{-1}=-\phi_{m2}.
	\label{}
\end{equation}
The gapping potential is modified accordingly to
\begin{equation}
	\mathcal{L}'=-u\cos (2\phi_l + \phi_{m1}-\phi_{m2}) +u \cos (2\phi_r + \phi_{m1}+\phi_{m2}).
	\label{}
\end{equation}
We now have
\begin{equation}
	\mathsf{R}W_s\mathsf{R}^{-1}=-W_s,
	\label{}
\end{equation}
implying $\eta_s=-1$.

If we ignore the auxilary Dirac fermion $\psi_{m2}$, the domain wall Dirac fermion $\psi_{m1}$ transforms as $\mathsf{R}\psi_{m1}\mathsf{R}^{-1}=-\psi_{m1}$. This is in fact the ``conjugate'' phase of the one discussed in the main text. To see why this is the case, we stack the two states together and the surface domain wall now has two Dirac fields $\psi_{m+}$ and $\psi_{m-}$ which transform as
\begin{equation}
	\mathsf{R}\psi_{m\pm}\mathsf{R}^{-1}=\pm\psi_{m\pm}.
	\label{}
\end{equation}
The two can be gapped out by introducing a pair of ``trivial'' chiral Dirac fields $\psi_l, \psi_r$ with opposite chirality, transforming under $\mathsf{R}$ as $\mathsf{R}\psi_{L}\mathsf{R}^{-1}=\psi_r$. All modes then can be gapped by the following coupling:
\begin{equation}
	\delta\psi_{m+}^\dag (\psi_r+\psi_l) +\delta \psi_{m-}^\dag (\psi_r-\psi_l) + \text{h.c.}
	\label{}
\end{equation}

\section{$\mathrm{Sp}(n)_n$ STO}
In this section we show that $\mathrm{Sp}(n)_n$ theory can arise as the STO for $\nu=2n$ TCSC. To this end we exploit the following conformal embedding
\begin{equation}
	\sp(n)_n \times \sp(n)_n \rightarrow \so(4n^2)_1.
	\label{}
\end{equation}

First, we explain how the two copies of matrix Lie algebra $\sp(n)$ embed into $\so(4n^2)$. We notice that $\mathrm{Sp}(n)_n$ can be viewed as the group of unitary $n\times n$ \emph{quaternion} matrices. A quaternion can be written as $a=a_0+\sum_{i=1}^3 a_i \vec{e}_i$ where $a_0, a_i\in \mathbb{R}$, and the $\vec{e}_i$'s satisfy $\vec{e}_i^2=-1$ and $\vec{e}_i\vec{e}_j=\epsilon_{ijk}\vec{e}_k$ for $i\neq j$. The multiplication of two quaternions $a$ and $b$ is given by
\begin{equation}
	a\cdot b=a_0b_0-a_ib_i + \sum_{k=1}^3 (a_0b_k+a_kb_0+\epsilon_{ijk}a_ib_j)\vec{e}_k.
	\label{}
\end{equation}
Conjugation is defined as $\bar{a}=a_0-\sum_{i=1}^3 a_i\vec{e}_i$. 
The Lie algebra $\sp(n)$ is given by $n\times n$ quaternionic matrices $B$ which satisfy
\begin{equation}
	B+\bar{B}^\mathsf{T}=0.
	\label{}
\end{equation}
In terms of components:
\begin{equation}
	B_{ab}^0=-B_{ba}^0, B_{ab}^i=B_{ba}^i.
	\label{}
\end{equation}
Therefore $\sp(n)$ has $\frac{n(n-1)}{2}+3\frac{n(n+1)}{2}=n(2n+1)$ generators.

We now discuss the embedding of $\sp(n)$ into $\so(4n^2)$. Consider a $n\times n$ quaternionic matrix $X$. We can mutiply a quaternionic matrix $B$ from $\sp(n)$ with $X$ from left or from right. Now we unpack $X$ to get a $4n^2$-dimensional real vector. The left/right multiplication of $B$ then gives two $4n^2\times 4n^2$ real skew symmetric matrices. 

Let us make this more explicit. For the left multiplication, we have
\begin{equation}
	(B\cdot X)_{ab}=B_{ac}^0 X_{cb}^0 - B_{ac}^i X_{cb}^i + \sum_k (B_{ac}^0X_{cb}^k+B_{ac}^kX_{cb}^0 + \epsilon^{ijk}B_{ac}^iX_{cb}^j)\vec{e}_\vk.
	\label{}
\end{equation}
We write it as $\tilde{X}^i_{a_1a_2}=(\Sigma^l)_{a_1a_2,b_1b_2}^{ij}X_{b_1b_2}^j$, where $(\Sigma^l)^{ij}_{a_1a_2,b_1b_2}=L^{ij}_{a_1b_1}\delta_{b_1b_2}$, and
\begin{equation}
	\begin{gathered}
		L_{ab}^{00}=B_{ab}^0, L_{ab}^{0i}=-B^i_{ab},\\
	L_{ab}^{i0}=B_{ab}^i, L_{ab}^{ii}=B_{ab}^0,\\
	L_{ab}^{ij}=-\epsilon^{ijk}B^k_{ab}\quad i\neq j.
	\end{gathered}
		\label{}
\end{equation}
Clearly $\Sigma^l$ is a skew-symmetric matrix.

Similarly we define $\Sigma^r$ from right multiplication: $(\Sigma^r)^{ij}_{a_1a_2,b_1b_2}=\delta_{a_1b_1}R^{ij}_{a_2b_2}$, where $R$ is defined by
\begin{equation}
	\begin{gathered}
		R_{ab}^{00}=B_{ab}^0, R_{ab}^{0i}=-B^i_{ab},\\
	R_{ab}^{i0}=B_{ab}^i, R_{ab}^{ii}=B_{ab}^0,\\
	R_{ab}^{ij}=\epsilon^{ijk}B^k_{ab}\quad i\neq j.
	\end{gathered}
		\label{}
\end{equation}
$L$ and $R$ only differ by a minus sign in the $ij$ entry, which turns out to be actually crucial.

Now consider $4n^2$ free chiral Majorana fermions $\chi_{ab}^i$, where $a,b=1,2,\dots,n$ and $i=0,1,2,3$. We define two $\sp(n)_n^{l/r}$ currents
\begin{equation}
	\mb{J}_{\sp(n)_n^{l/r}}=\frac{i}{2}\chi^\mathsf{T}\bm{\Sigma}^{l/r}\chi.
	\label{}
\end{equation}
Here $\bm{\Sigma}^{l/r}$ are the set of matrices derived from generators of $\sp(n)$. 
It is straightforward to show that: (1) The currents we define indeed satisfy the current algebra of $\sp(n)_n$. (2) The energy-momentum tensor of $\so(4n^2)_1$ decomposes.
The Hamiltonian is a straightforward generalization of the one for $\mathrm{SO}(\nu)_\nu$ in the main text.

Now we need to impose reflection symmetry on the Majorana fields, such that the reflection maps the two subalgebras into each other. Clearly we need $\chi_{a_1a_2}\leftrightarrow \chi_{a_2a_1}$. However, this is not enough: due to the minus sign in the $ij$ entry, we need an additional exchange of two indices among $1,2$ and $3$. Which pair to choose does not matter, say $1$ and $2$. Therefore, the reflection-invariant Majorana modes are $\chi_{aa}^0$ and $\chi_{aa}^3$ for $a=1,2,\dots,n$ and there are $2n$ of them. We thus conclude that the bulk has the topological invariant $\nu=2n$.

\section{STOs of Topological Crystalline Insulators}
Topological crystalline insulators in 3D are classified by the mirror Chern number $n\in\mathbb{Z}$. A reflection-symmetric surface hosts $\nu$ Dirac cones:
\begin{equation}
\mathcal{H}_0=\sum_{a=1}^{n} \psi_a^\dag ( -i\sigma_y\partial_x +i\sigma_x\partial_y) \psi_a,
	\label{}
\end{equation}
where reflection acts on the Dirac fields as
\begin{equation}
	\mathsf{R}\psi_a(x,y)\mathsf{R}^{-1}=\sigma_x\psi_a(-x,y).
	\label{}
\end{equation}
 Again we can turn on the following spatially-varying mass:
\begin{equation}
	H_1=\int\di^2\vr\, M(x)\psi^\dag_a(\vr)\sigma_z\psi_a(\vr),
	\label{}
\end{equation}
in which $M(x)$ is an odd function of $x$: $M(-x)=-M(x)$ to preserve the reflection symmetry. We find that there are $\nu$ chiral Dirac fermions living on the domain wall.

\subsection{$n=1$: T-Pfaffian }
We first construct a charge-conserving non-Abelian STO for the $\nu=1$ TCI, the T-Pfaffian state~\cite{Bonderson13d, chen2014b, metlitski2014}. First we briefly review the topological order of T-Pfaffian. It can be described by a quotient theory $[\text{Ising}\times\mathrm{U}(1)_{-8}]/\mathbb{Z}_2$: the quasiparticles are labeled by $a_j$ where $a=I, \sigma, \psi$ labels anyon types in the Ising theory, and $[j]$ labels anyon types in the $\mathrm{U}(1)_{-8}$ theory, subject to the rule that odd $j$'s have to pair up with $a=\sigma$ and even $j$'s  can  pair up with $a=I, \psi$. $\psi_4$ has trivial braiding with all other particles, and is identified as the physical fermion.

The edge theory is described by a neutral chiral Majorana field and a charged chiral boson:
\begin{equation}
	\mathcal{L}=i\chi(\partial_t+v_n\partial_y)\chi-\frac{2}{4\pi}\partial_t\phi\partial_y\phi - v_c (\partial_y\phi)^2.
	\label{}
\end{equation}
Here $v_n, v_c >0$ set the velocities of the two modes. Importantly, the edge theory posseses a $\mathbb{Z}_2$ gauge symmetry generated by
\begin{equation}
	\chi\rightarrow -\chi, \phi\rightarrow \phi+\frac{\pi}{2},
	\label{}
\end{equation}
which guarantees that local operators are built out of electron operators $\chi e^{-2i\phi}$. Under a $\mathrm{U}(1)$ phase rotation of angle $\alpha$, $\phi$ transforms as $\phi\rightarrow \phi-\frac{\alpha}{2}$.
We denote this edge theory as $[\text{MF}\times \mathrm{U}(1)_{-2}]/\mathbb{Z}_2$. 

We turn on the following gapping potential:
\begin{equation}
	\mathcal{L}_1=-u\cos 4(\phi_l+\phi_r+\phi_m).
	\label{}
\end{equation}
Schematically, this term can be represented in terms of electron operators as $\psi_l^2\psi_r^2 (\psi_m^\dag)^4+\text{h.c.}$.  Naively the potential has multiple minima corresponding to $\langle \phi_l+\phi_r+\phi_m\rangle=\frac{k\pi}{2}$ where $k\in \mathbb{Z}$. However, they are in fact related to each other via the $\mathbb{Z}_2$ gauge transformation $\phi_l\rightarrow \phi_l+\frac{\pi}{2}$, so there is only a unique ground state. Namely, the would-be ``order parameter'' $e^{2i(\phi_{L}+\phi_r+\phi_m)}$ is not a local operator. We then choose a gauge in which $\langle \phi_l+\phi_r+\phi_m\rangle=0$. The term essentially identifies the $\mathbb{Z}_2$ gauge symmetries from the two theories.

Let us determine what are the remaining fields. The Majorana fermions $\chi_l$ and $\chi_r$ are not affected by the gapping potential. For the three bosons $\phi_l, \phi_r$ and $\phi_m$, the gapping potential freezes the combination $\phi_l+\phi_r+\phi_m$, therefore any linear combination that does not commute with it must be gapped. The remaining propagating field is $\tilde{\phi}_m = 2\phi_l + \phi_m\simeq -2\phi_r -\phi_m$. One can easily check that $\tilde{\phi}_m$ satisfies a commutation relation $[\tilde{\phi}_m(y), \partial_{y'}\tilde{\phi}_m(y')]=i\pi\delta(y-y')$, which means that we can refermionize $\tilde{\psi}_m=e^{-i\tilde{\phi}_m}$. Notice that $\tilde{\psi}$ is not gauge-invariant: it transforms as $\tilde{\psi}_m\rightarrow -\tilde{\psi}_m$. While under the reflection, we still have
\begin{equation}
	\mathsf{R}(\tilde{\psi}_m)=\tilde{\psi}_m^\dag.
	\label{}
\end{equation}

We then add the following electron tunneling terms and project to the low-energy subspace:
\begin{equation}
	\begin{split}
		\mathcal{L}_2&=t(\psi_l^\dag +\psi_r^\dag) \psi_m + \text{h.c.}\\
		&\simeq t\chi_l \tilde{\psi}_m^\dag +  t\chi_r \tilde{\psi}_m+\text{h.c.}\\
		&\sim i\Im(t) \chi_+\chi_1-i\Re (t) \chi_- \chi_2,
	\end{split}
	\label{}
\end{equation}
where we have introduced $\chi_\pm = \frac{\chi_l\pm\chi_r}{\sqrt{2}}, \tilde{\psi}_m=\frac{\chi_1+i\chi_2}{2}$. Notice that $\tilde{\psi}_m$ and $\chi_{L/R}$ have opposite chiralities. When $t$ is \emph{complex}, all remaining fermions are gapped out while preserving the reflection symmetry.

\subsection{$n=2$: $\mathbb{Z}_4$ gauge theory}
\label{sec:Z4}
For $n=2$, there are two chiral Dirac fermions on the domain wall, denoted by $\psi_{mj}$ with $j=1,2$. We first add an additional pair of chiral Dirac fermions $\psi_l$ and $\psi_r$ propagating in the opposite direction. Under reflection they transform into each other:
$\mathsf{R}\psi_l\mathsf{R}^{-1}=\psi_r$.
Therefore, the addition does not change the bulk and can be considered as a kind of surface reconstruction. We can form the linear combinations $\psi_\pm=\frac{1}{\sqrt{2}}(\psi_l\pm \psi_r)$, and turn on a coupling term $-\delta (\psi_+^\dag \psi_{m2}+\text{h.c.})$ to gap out a pair of counter-propagating Dirac fermions. We are left with non-chiral Dirac fermions $\psi_{m1}$ and $\psi_-$, with the reflection action given by
\begin{equation}
	\mathsf{R}\psi_{m1}\mathsf{R}^{-1}=\psi_{m1}, \mathsf{R}\psi_-\mathsf{R}^{-1}=-\psi_-.
	\label{}
\end{equation}
They can be bosonized $\psi_{m1}\sim e^{i\phi_{m1}}, \psi_-\sim e^{-i\phi_{m2}}$, and under reflection $\mathsf{R}\phi_{m2}\mathsf{R}^{-1}=\phi_{m2}+\pi$.

To gap out the edge modes, we introduce a $\mathbb{Z}_4$ toric code on the surface.  The low-energy Luttinger liquid theory is defined by the following K matrix:
\begin{equation}
	\mathcal{K}=
	\begin{pmatrix}
		4\sigma^x & 0 & 0\\
		0 & 4\sigma^x & 0\\
		0 & 0 & \sigma^z
	\end{pmatrix},
	\label{}
\end{equation}
with the bosonic fields being $\Phi=(\phi_{L1}, \phi_{L2}, \phi_{R1}, \phi_{R2}, \phi_{m1}, \phi_{m2})^\mathsf{T}$. We postulate that the charge vector takes the form $\mb{t}=(0,2,0,2,1,1)$. Namely, $e$ anyons (represented by $e^{i\phi_{L1}}$ or $e^{i\phi_{R1}}$) in the $\mathbb{Z}_4$ toric code carry half electric charges. We note that although in principle an $e$ anyon can carry $1/4$ electric charge, this is forbidden by the charge-statistics relations for a system of electrons: since $e^4$ is a bosonic local excitation, it has to carry even number of electric charges.

We add the following gapping potentials:
\begin{equation}
	\begin{split}
\mathcal{L}'=& - u\cos 4(\phi_{L1}+\phi_{R1}+\phi_{L2}+\phi_{R2}+\phi_{m2})
-v\cos (4\phi_{L1} - \phi_{m1}+\phi_{m2}) 
+ v\cos (4\phi_{R1}- \phi_{m1} + \phi_{m2}). \\
	\end{split}
	\label{}
\end{equation}
It is not difficult to see that the gapping potentials result in a fully gapped state with a unique ground state.

Let us determine how anyons are permuted by the reflection. Since $\phi_{L1}+\phi_{R1}+\phi_{L2}+\phi_{R2}+\phi_{m2}$ is pinned, we have
\begin{equation}
	\mathsf{R}(em)= \bar{e}\bar{m}f. 
	\label{eqn:R-em}
\end{equation}
We can also see that $\eta_{em}=-1$.  

The last two terms imply that $\phi_{L1}-\phi_{R1}$ is pinned, so we find
\begin{equation}
	\mathsf{R}(e)=e.
	\label{eqn:R-e}
\end{equation}
In addition, the last two terms imply that $\langle 2\phi_{L1}-2\phi_{R1}\rangle=\pm\frac{\pi}{2}$, which leads to $\eta_{e^2}=-1$ following a calculation similar to the one in Sec. \ref{sec:toriccode}.

Combining Eqs. \eqref{eqn:R-em} and \eqref{eqn:R-e} we find that 
\begin{equation}
	\mathsf{R}(m)= e^2\bar{m}f.
	\label{}
\end{equation}
 
As a consistency check, we also find the following superposition of pinned fields:
\begin{equation}
4(\phi_{L1}+\phi_{R1}+\phi_{L2}+\phi_{R2}+\phi_{m2}) -
(4\phi_{L1} - \phi_{m1}+\phi_{m2}) - (4\phi_{R1}- \phi_{m1} + \phi_{m2}) =2 (2\phi_{L2}+2\phi_{R2}+\phi_{m1}+\phi_{m2}), 
	\label{}
\end{equation}
which implies $\eta_{m^2}=-1$. This is consistent with $\eta_{e^2}\eta_{m^2}=\eta_{(em)^2}=1$.

The STO can be simplified if we condense the charge-$1$ anyon $e^2m^2$, which breaks the charge-conservation symmetry. In the condensed phase there are four remaining anyons: $1, e^2\sim m^2, em\sim \bar{e}\bar{m}, \bar{e}m\sim e\bar{m}$, which is the double semion topological order. Notice that different from double semion realized in two dimensions, here the boson $e^2$ has $\eta_{e^2}=-1$, which is impossible in two dimensions. In fact, the theory is nothing but two copies of semion-fermion theories.

\section{Decomposition of the Energy-Momentum Tensor in the Conformal Embedding}
\def\son{\mathfrak{so}(\nu)_\nu}

The Sugawara energy-momentum tensors are given by the following normal ordered product:
\begin{equation}
	T_{\son^\lambda}(z)=\frac{1}{8(\nu-1)}\mb{J}_{\lambda}(z)\cdot \mb{J}_{\lambda}(z), \lambda=l,r.
	\label{}
\end{equation}
We need the following OPE relation
\begin{equation}
	\chi_a\chi_b\chi_a\chi_b=\chi_a\partial\chi_a + \chi_b\partial\chi_b.
	\label{}
\end{equation}

Let us first calculate the explicit forms of $\mb{J}_{\pm}$:
\begin{equation}
	\begin{gathered}
	J_{l}^{(r,s)}(z)=i\sum_{a=1}^\nu \chi_{ra}(z)\chi_{sa}(z),\\
	J_{r}^{(r,s)}(z)=i\sum_{a=1}^\nu \chi_{ar}(z)\chi_{as}(z).
	\end{gathered}
	\label{}
\end{equation}

We can then easily derive
\begin{equation}
	\begin{split}
	\sum_{r<s}J_{l}^{(r,s)}\cdot J_{l}^{(r,s)} &=-\sum_{r<s}\left(\sum_{a=1}^\nu \chi_{ra}\chi_{sa}\chi_{ra}\chi_{sa} + \sum_{a\neq b}\chi_{ra}\chi_{sa}\chi_{rb}\chi_{sb}\right)\\	
	&=-\sum_{r<s}\left(\sum_a (\chi_{ra}\partial\chi_{ra}+\chi_{sa}\partial\chi_{sa})+2\sum_{a<b}\chi_{ra}\chi_{sa}\chi_{rb}\chi_{sb}\right)\\
	&=-2(\nu-1)\sum_{r,a}\chi_{ra}\partial\chi_{ra} -2\sum_{r<s}\sum_{a<b}\chi_{ra}\chi_{sa}\chi_{rb}\chi_{sb}.
	\end{split}
	\label{}
\end{equation}
Similarly we find
\begin{equation}
	\begin{split}
	\sum_{r<s}J_{r}^{(r,s)}\cdot J_{r}^{(r,s)} =-2(\nu-1)\sum_{r,a}\chi_{ra}\partial\chi_{ra} -2\sum_{r<s}\sum_{a<b}\chi_{ar}\chi_{as}\chi_{br}\chi_{bs}.
	\end{split}
	\label{}
\end{equation}
We will relabel the dummy indices:
\begin{equation}
	\sum_{r<s}\sum_{a<b}\chi_{ar}\chi_{as}\chi_{br}\chi_{bs}=\sum_{a<b}\sum_{r<s}\chi_{ra}\chi_{rb}\chi_{sa}\chi_{sb}.
	\label{}
\end{equation}
Therefore when we add up the energy-momentum tensors from $\son^l$ and $\son^r$, the four-fermion terms exactly cancel and we have
\begin{equation}
	T_{\son^l}+T_{\son^r}=-\frac{1}{2}\sum_{r,a}\chi_{ra}\partial\chi_{ra}.
	\label{}
\end{equation}

\twocolumngrid
\bibliography{TI}

\end{document}